\documentclass[aps,prl,reprint,superscriptaddress]{revtex4-1}
\usepackage[separate-uncertainty=true,multi-part-units=single]{siunitx}
\usepackage{graphicx,amsmath}
\usepackage{bm}

\begin{document}

%\preprint{}

\title{Negative spin-Hall angle and anisotropic spin-orbit torques in epitaxial IrMn}


\author{V.~Tshitoyan}
\email[]{vahe.tshitoyan@gmail.com}
%\thanks{}
\altaffiliation{Present address: Zedsen Ltd., 3 Chiswick Business Park, 566 Chiswick High Road, London, W4 5YA, UK}
\affiliation{Microelectronics Group, Cavendish Laboratory, University of Cambridge, CB3 0HE, UK}
\author{P.~Wadley}
\affiliation{School of Physics and Astronomy, University of Nottingham, Nottingham, NG7 2RD}
\author{M.~Wang}
\affiliation{School of Physics and Astronomy, University of Nottingham, Nottingham, NG7 2RD}
\author{A.~W.~Rushforth}
\affiliation{School of Physics and Astronomy, University of Nottingham, Nottingham, NG7 2RD}
\author{A.~J.~Ferguson}
\altaffiliation{Present address: Evonetix Ltd., Chesterford Research Park, Cambridge, CB10 1XL, UK.}
\affiliation{Microelectronics Group, Cavendish Laboratory, University of Cambridge, CB3 0HE, UK}


\date{\today}

\begin{abstract}
A spin-torque ferromagnetic resonance study is performed in epitaxial Fe / $\mathrm{Ir_{15}Mn_{85}}$ bilayers with different Fe thicknesses. We measure a negative spin-Hall angle of a few percent in the antiferromagnetic IrMn in contrast to previously reported positive values. A large spin-orbit field with Rashba symmetry opposing the Oersted field is also present. Magnitudes of measured spin-orbit torques depend on the crystallographic direction of current and are correlated with the exchange bias direction set during growth. We suggest that the uncompensated moments at the Fe / IrMn interface are responsible for the observed anisotropy. Our findings highlight the importance of crystalline and magnetic structures for the spin-Hall effect in antiferromagnets.
\end{abstract}

%\keywords{}

\maketitle

%\section{\label{sec:introduction}Introduction}
Following recent breakthroughs in antiferromagnetic spintronics~\cite{Jungwirth2016,MacDonald2011} the potential of antiferromagnets in electrical and information technology is being widely investigated. Several studies have achieved electrical manipulation of antiferromagnetic moments~\cite{Wei2007x,Tang2007,Urazhdin2007,Moriyama2015a,Wadley2016x}, transfer of spin-polarization across large distances through antiferromagnets~\cite{Wang2014,Hahn2014,Moriyama2015}, as well as efficient manipulation of ferromagnets using antiferromagnets~\cite{Tshitoyan2015,Zhang2015,Reichlova2015}, leading to magnetic field-free switching of ferromagnets~\cite{VandenBrink2016,Kong2016,Oh2016,Fukami2016}. Magnetoresistance effects have been measured in several antiferromagnets~\cite{Park2011a,Wang2012a,Marti2014,Fina2014,Kriegner2016}. Many of these findings come together in a recent study demonstrating an antiferromagnetic memory device~\cite{Wadley2016x}. There, electrical current in the antiferromagnet induces locally varying spin-polarization due to spin-orbit fields, which is able to switch the staggered antiferromagnetic moments at sufficiently large current densities. The switching is measured using the anisotropic magnetoresistance of the antiferromaget. 

Several studies have focused on the spin-Hall effect in antiferromagnets~\cite{Zelezny2017,Zhang2017,Zhang2014,Mendes2014}. It is predicted that in non-collinear antiferromagnets such as IrMn the spin-Hall angle depends on the specific arrangement of magnetic moments with respect to the current. By switching between different magnetic arrangements one can potentially achieve new device functionalities, such as a tunable spin-current source. Here, we report the first measurement of a negative spin-Hall angle in antiferromagnetic IrMn, a material for which previously only a positive spin-Hall angle has been reported~\cite{Tshitoyan2015,Reichlova2015,Zhang2015,VandenBrink2016,Kong2016,Oh2016,Zhang2014,Mendes2014,Wu2016,Zhang2016}.

\begin{figure}[b!]
	\includegraphics[width=1.0\columnwidth,angle=0]{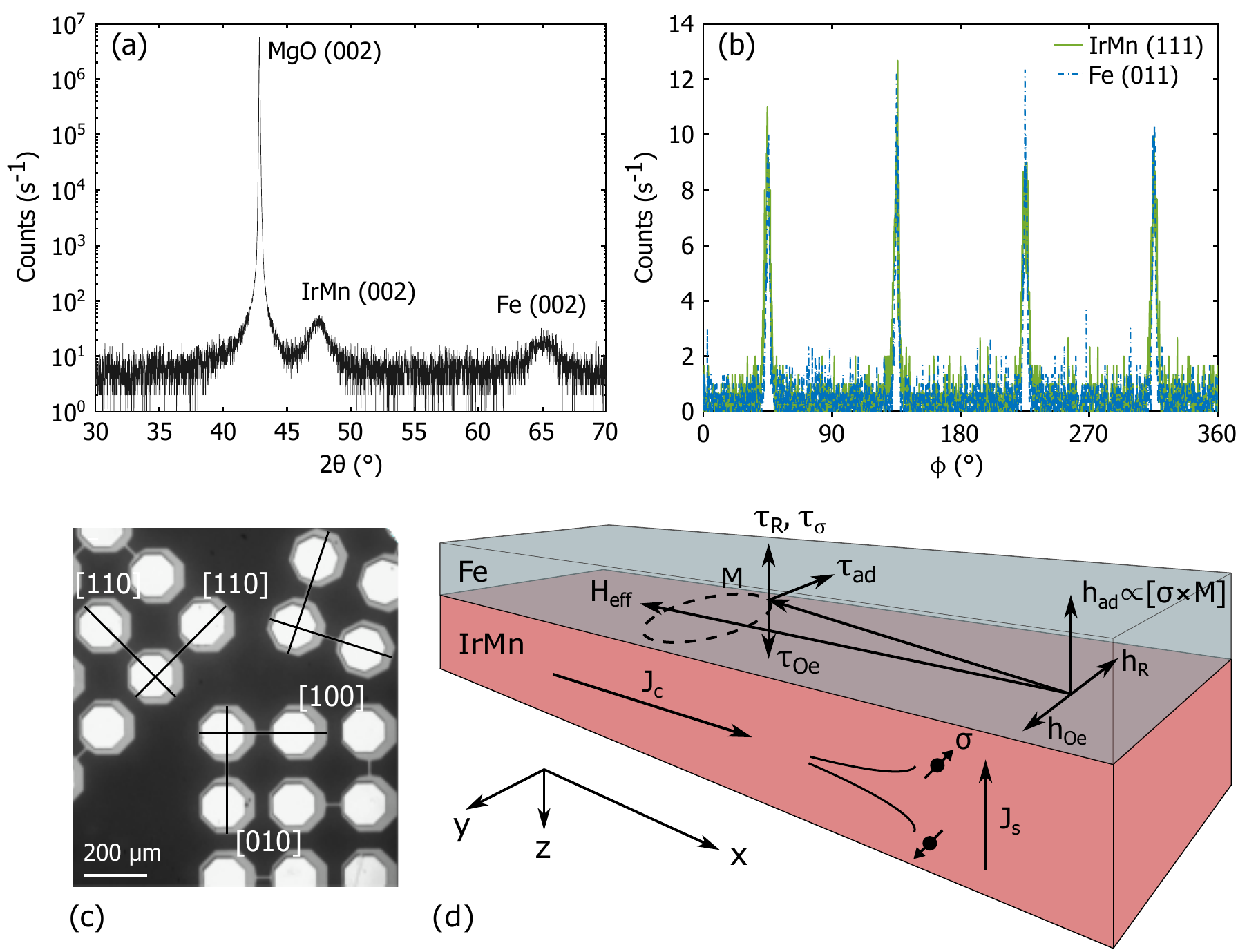}
	\caption{(a, b) X-ray diffraction measurements of MgO / Fe(10) / $\mathrm{Ir_{15}Mn_{85}}$(20) / Ta(3) film grown in the same conditions as the thinner films used in this study. (a) 2$\theta$ rotation showing the out-of-plane crystalline order. The angle of incidence of the X-ray beam is $\pi/2 - \theta$. The peaks correspond to the (002) lattice spacings of the individual layers. (b) In-plane $\phi$ rotations at two different $\theta$ corresponding to IrMn~(111) and Fe~(011). The overlapping peaks for (111) and (011) directions indicate an in-plane $\SI{45}{\degree}$ orientation between the Fe and the IrMn crystals, as expected from their lattice parameters of $\SI{2.87}{\angstrom}$ and $\SI{3.78}{\angstrom}$~\cite{Yamaoka1974,Sakuma2003}. (c) An optical micrograph of the bars aligned with different crystallographic directions of Fe. (d) Microwave charge current $\mathbf{J_c}$ induces torques driving the precession of the Fe magnetization $\mathbf{M}$ around $\mathbf{H_{eff}}$. $\mathbf{J_s}$ is the spin current, $\bm{\sigma}$ is the spin-Hall polarization, $\bm{\tau}$ and  $\mathbf{h}$ are the various current induced torques and corresponding effective fields.}
	\label{fig:samples}
\end{figure}
%\section{\label{sec:materials}Materials}
MgO / Fe(2, 3, 4) / $\mathrm{Ir_{15}Mn_{85}}$(5) / Al(2) multilayers are sputtered on an MgO(001) substrate annealed at $\SI{500}{\degreeCelsius}$ and $\mathrm{P}<\SI{e-8}{Torr}$ before the deposition. Numbers in parentheses are layer thicknesses in nanometers. Composition of the IrMn is determined from thicknesses of Ir and Mn calibration films measured using X-ray reflectometry. The Fe and the IrMn are grown at $\SI{195}{\degreeCelsius}$. The protective Al layer is grown at $\SI{30}{\degreeCelsius}$. Typical growth rates are 0.1 - $\SI{0.5}{\angstrom\per\second}$. The elevated deposition temperature and the small rate lead to epitaxial growth of Fe and IrMn, facilitated by the matching lattice parameter of the MgO substrate. X-ray diffraction is used to confirm the out-of-plane (001) order with a  2$\theta$-rotation (Fig.~\ref{fig:samples}(a)), whereas the in-plane four-fold symmetry is apparent from $\phi$-rotations (Fig.~\ref{fig:samples}(b)). Magnetic field of a few tens of mT is applied during growth to set the exchange bias. The measured exchange biases vary from $10$ to $\SI{30}{\milli\tesla}$ for different Fe thicknesses, confirming the antiferromagnetic order in all films.  

%\section{\label{sec:measurement-technique}Measurement Technique}
Bars of $\SI{5}{\micro\meter}$ width and $\SI{45}{\micro\meter}$ length are fabricated using photolithography and ion milling (Fig.~\ref{fig:samples}(c)). Contact pads of  Cr(7) / Au(70) are thermally evaporated for wire bonding. One end of the bar is connected to a microwave transmission line, whereas the other end is grounded. Applied microwave current induces torques that drive the precession of the magnetization of Fe around the saturating magnetic field $\mathbf{H_{eff}}$ (Fig.~\ref{fig:samples}(d)). Several field-like and (anti)damping-like torques can coexist in such bilayers. Examples of out-of-plane field-like torques are the Oersted torque $\bm{\tau}_\mathbf{Oe}\propto-[\mathbf{h_{Oe}}\times\mathbf{M}]$, the interfacial Rashba spin-orbit torque $\bm{\tau}_\mathbf{R}\propto-[\mathbf{h_R}\times\mathbf{M}]$~\cite{Miron2010,Miron2011} and the field-like component of the spin-transfer torque $\bm{\tau_\sigma}\propto-[\bm{\sigma}\times\mathbf{M}]$ due to the spin-Hall current from IrMn~\cite{Liu2012}. An in-plane (anti)damping-like component of the spin-transfer torque given by $\bm{\tau}_\mathbf{ad}\propto\mathbf{M}\times[\bm{\sigma}\times\mathbf{M}]$ is also present in bilayers with heavy metals, and can be described by an out-of-plane field 
\begin{equation}
\label{eqn:had}
\mathbf{h_{ad}}\propto[\bm{\sigma}\times\mathbf{M}] \propto \mathbf{\hat{z}}h_{ad}\cos\phi.
\end{equation}

We measure the current-induced magnetization precession as a rectified direct voltage due to the anisotropic magnetoresistance (AMR)~\cite{Tulapurkar2005x,Costache2006,Liu2011x}. Microwave frequency is kept constant while the magnetic field is varied in the plane of the film. The measured voltage can be decomposed into symmetric and antisymmetric Lorentzians centered at the ferromagnetic resonance field $H_0$, so that
\begin{equation}
\label{eqn:vdc-spin-torque-fmr}
V_{dc} = V_{sym}\frac{\Delta{H}^2}{(H-H_0)^2+\Delta{H}^2} + V_{asy}\frac{\Delta{H}(H-H_0)}{(H-H_0)^2+\Delta{H}^2},
\end{equation}
\noindent with magnitudes given by
\begin{equation}
\label{eqn:epi-irmn-vsym-vasy}
\begin{aligned}
V_{sym} &= -\frac{J\Delta{R}}{2}A_{sym}\sin 2\phi \cdot h_z, \\
V_{asy} &= -\frac{J\Delta{R}}{2}A_{asy}\sin 2\phi \cdot (h_y \cos\phi - h_x\sin\phi), \\
\end{aligned}
\end{equation}
where $J$ is the current in the bar, $\Delta{R}$ is the AMR amplitude, $A_{sym}$ and $A_{asy}$ are coefficients determined by the magnetic anisotropies, and $\phi$ is the angle between the magnetization and current directions~\cite{TshitoyanThesis}. The current-induced field $\mathbf{h}(h_x, h_y, h_z)$ is a combination of the earlier discussed effective fields shown in Fig.~\ref{fig:samples}(d).

\begin{figure}[t!]
	\includegraphics[width=1\columnwidth,angle=0]{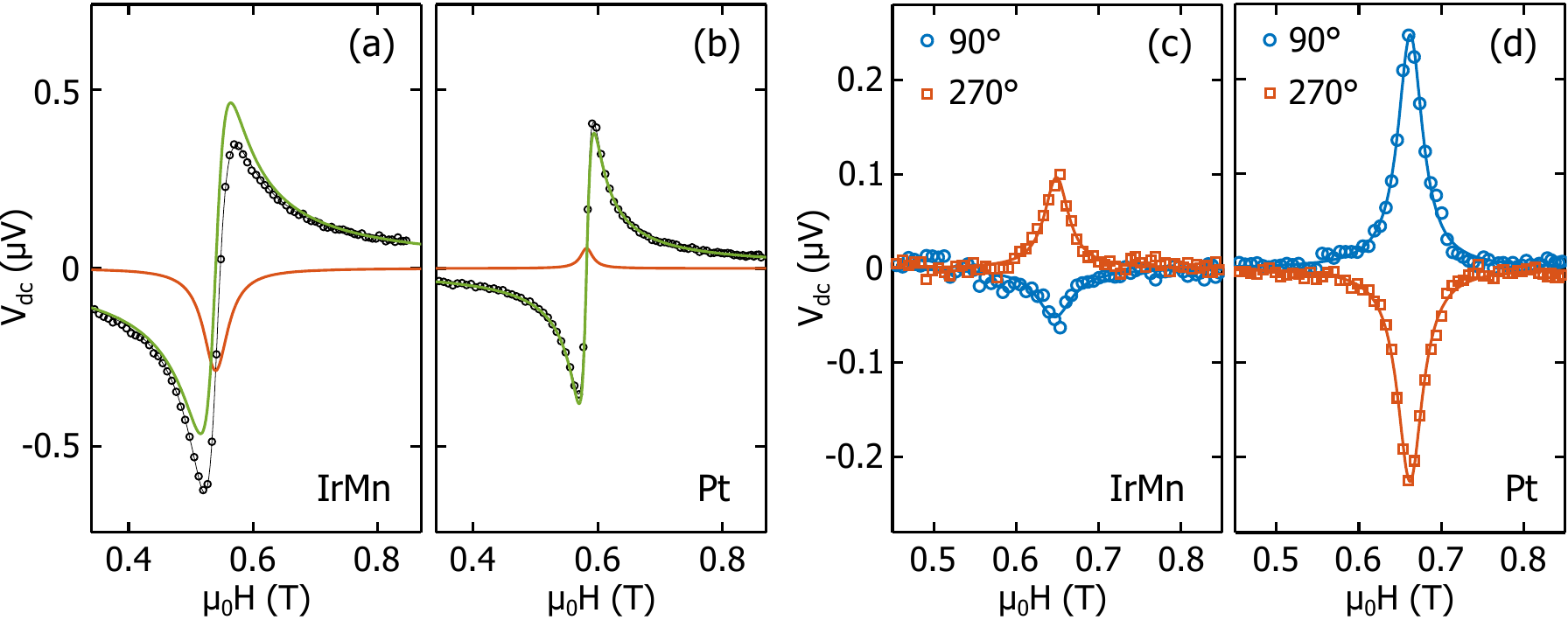}
	\caption{(a,~b) Spin-torque ferromagnetic resonance measured at $\SI{36.6}{\GHz}$ and $\phi=\SI{195}{\degree}$ in Fe(4) / IrMn(5) and Fe(4) / Pt(5) bars, both along the [100] crystallographic direction of Fe. Since the resonance field is large compared to the in-plane anisotropy and the exchange bias, the magnetization is assumed to be aligned with the external field. (c,~d) Spin-pumping measurements in Fe(4) / IrMn(5) and Fe(4) / Pt(5) bars for two opposite directions of magnetization perpendicular to the bar. The microwave field is along the bar. The solid lines are fits to symmetric Lorentzians. A small antisymmetric component is subtracted from the data. Frequencies of $\SI{36.3}{\GHz}$ (c) and $\SI{37.5}{\GHz}$ (d) are used to maximize the signal in each measurement. The sign of the signal does not depend on the frequency in the evaluated range of 27.5 to $\SI{37.5}{\GHz}$.}
	\label{fig:epi-irmn-negative-spin-hall-angle}
\end{figure}

%\section{\label{sec:negative-spin-hall-angle}Negative Spin-Hall Angle}
Ferromagnetic resonance measured in the Fe(4) / IrMn sample at $\SI{36.6}{\GHz}$ is fitted to  Eqn.~\ref{eqn:vdc-spin-torque-fmr} as shown in Fig.~\ref{fig:epi-irmn-negative-spin-hall-angle}(a). For the used angle of $\phi=\SI{195}{\degree}$, the positive $V_{asy}$ means $h_y>0$, because $\sin(2\phi)\cos(\phi)<0$ and $h_x=0$ (Eqn.~\ref{eqn:epi-irmn-vsym-vasy}). This is consistent with the Oersted field induced by the current in the IrMn (Fig.~\ref{fig:samples}(d)). The negative sign of $V_{sym}$ corresponds to $h_{ad}<0$ because $h_z = h_{ad}\cos\phi$. Therefore
\begin{equation}
\theta_{SH} = h_{ad}\frac{2e\mu_0M_sd_{Fe}}{\hbar J}<0,
\label{eqn:theta-sh}
\end{equation}
in contrast to previous measurements of positive spin-Hall angles in IrMn. Here, $d_{Fe}$ is the Fe thickness and $M_s$ is its saturation magnetization. The same measurement is repeated in a Fe(4)~/~Pt control sample and a positive symmetric component is observed (Fig.~\ref{fig:epi-irmn-negative-spin-hall-angle}(b)), as expected for the positive spin-Hall angle of Pt. 

To confirm the opposite signs of the spin-Hall angles of Pt and the measured IrMn we perform spin-pumping measurements~\cite{Saitoh2006,Mosendz2010,Tserkovnyak2005}. The chip is placed on a microstrip transmission line and voltage is measured across a bar perpendicular to the line. The microwave field is in-plane along the bar, and the external field is applied perpendicular to the bar. Opposite signs of inverse spin-Hall voltages are measured in the IrMn and Pt samples (Fig.~\ref{fig:epi-irmn-negative-spin-hall-angle}(c,~d)), confirming the opposite spin-Hall angles.

\begin{figure}[t!]
	\includegraphics[width=1\columnwidth,angle=0]{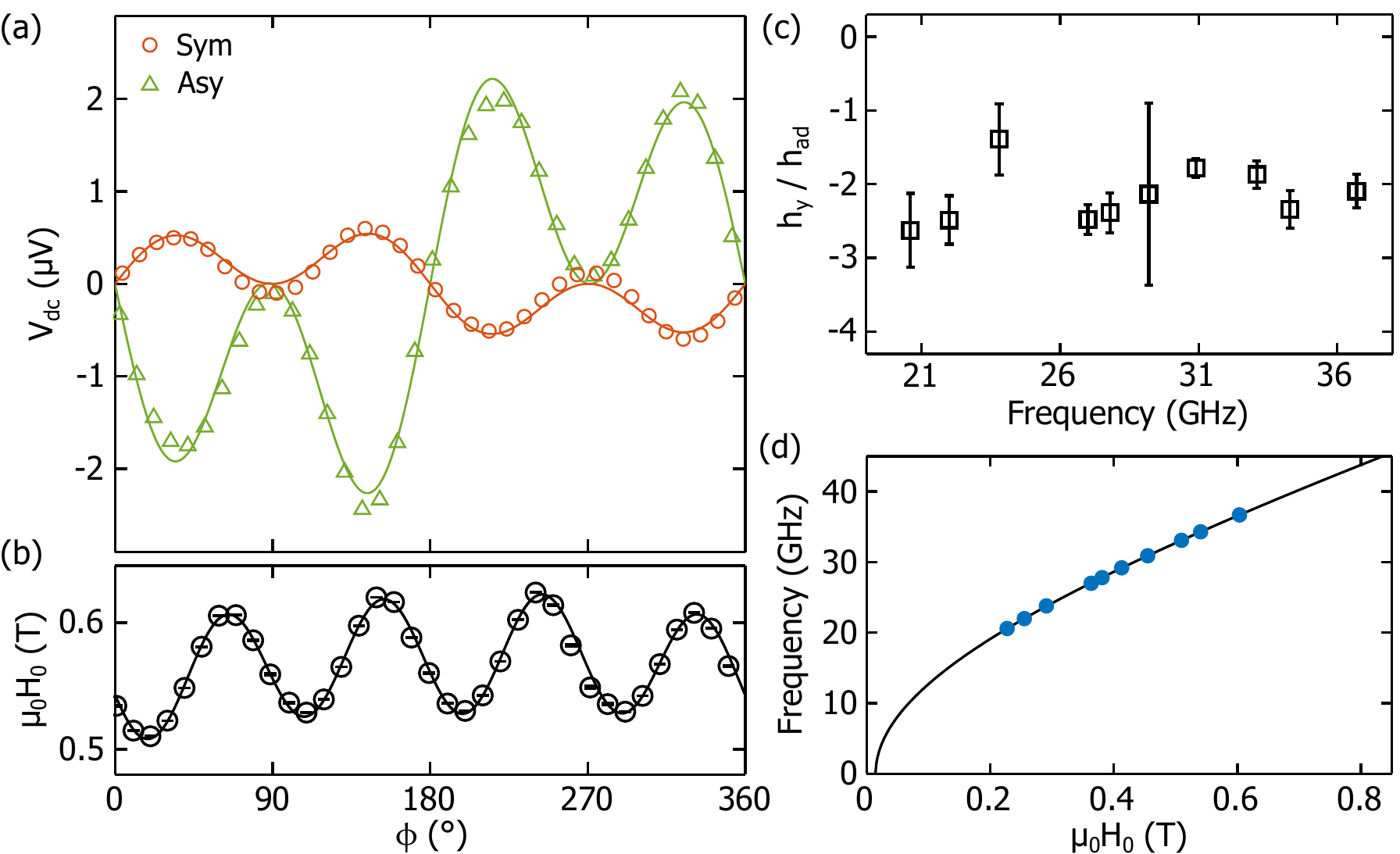}
	\caption{(a) In-plane angle-dependence of $V_{sym}$ and $V_{asy}$ in one of the Fe(4) / IrMn devices fitted to Eqn.~\ref{eqn:epi-irmn-vsym-vasy} using $h_z=h_{ad}\cos\phi+const$. (b) Angle dependence of the resonance frequency. The four-fold symmetry is due to the $\SI{52}{\milli\tesla}$ cubic anisotropy of Fe. There is also an exchange bias of $\SI{10}{\milli\tesla}$ at approximately $\SI{20}{\degree}$. (c) Frequency dependence of the $h_y/h_{ad}$ extracted from FMR measurements at $\phi=\SI{55}{\degree}$, showing no systematic variation in a large frequency range. (d) Microwave frequency and the corresponding resonance field fitted to Kittel's equation~\cite{Kittel1948}. The error bars are smaller than the size of the data points. $M_{eff}=\SI{2.32\pm0.02}{\tesla}$ is extracted from the fit.}
	\label{fig:epi-irmn-symmetries-kittel}
\end{figure}

$V_{sym}$ and $V_{asy}$ for different angles $\phi$ in one of the Fe(4)~/~IrMn devices are plotted in Fig.~\ref{fig:epi-irmn-symmetries-kittel}(a). The symmetries are consistent with $h_z\propto\cos\phi$ and an in-plane constant field $h_y$, and are found to be independent on the crystallographic direction of the current. Strong cubic anisotropy of Fe as well as an exchange bias are measured (Fig.~\ref{fig:epi-irmn-symmetries-kittel}(b)), confirming the antiferromagnetic order in IrMn. The negative $h_y/h_{ad}$ ratio shows no systematic variation across the studied frequency range (Fig.~\ref{fig:epi-irmn-symmetries-kittel}(c)) and the resonance frequency is described well by the Kittel mode with an effective magnetization of $M_{eff}=\SI{2.32\pm0.02}{\tesla}$ (Fig.~\ref{fig:epi-irmn-symmetries-kittel}(d)). This is slightly larger than the saturation magnetization of $\SI{2.18}{\tesla}$ of bulk Fe~\cite{Bayreuther1983}, the value also measured in the Fe(4)~/~Pt control sample. The large value of $M_{eff}$ could be due to a partial polarization of the uncompensated IrMn moments at the Fe / IrMn interface.

\begin{figure}[b!]
	\includegraphics[width=1\columnwidth,angle=0]{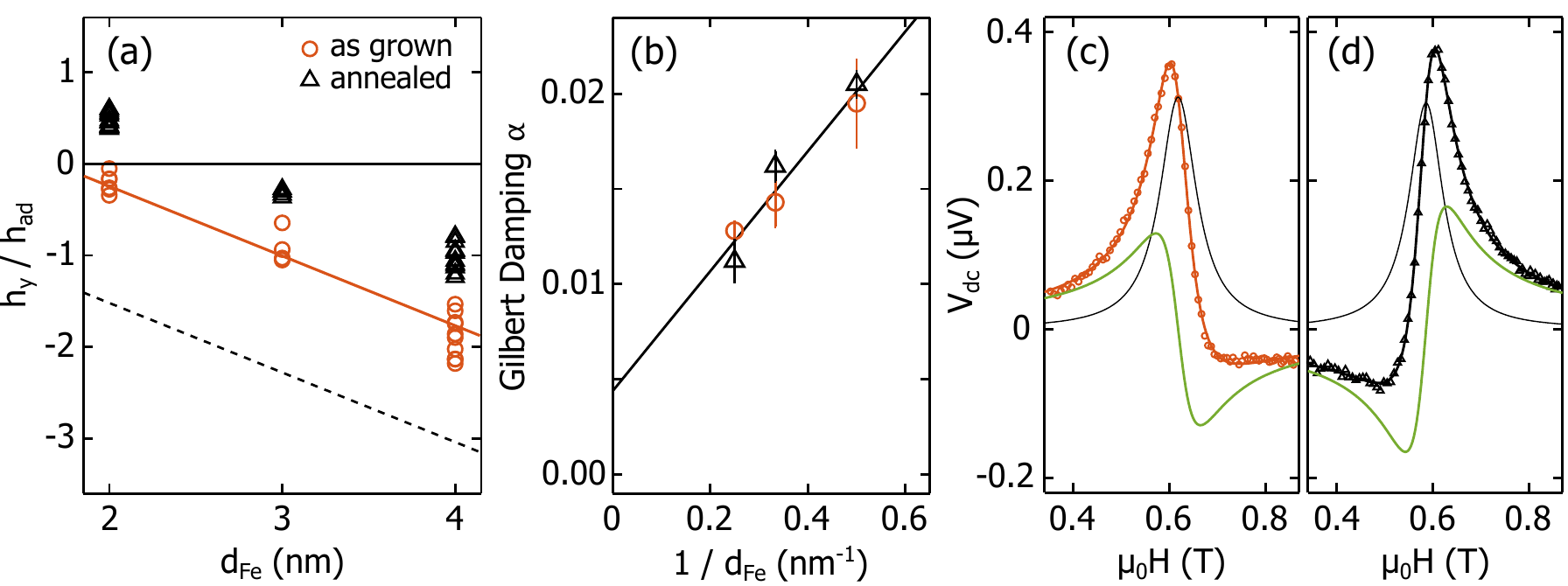}
	\caption{(a) $h_y/h_{ad}$ extracted from the in-plane rotations (Fig.~\ref{fig:epi-irmn-symmetries-kittel}(a)) for multiple devices for each Fe thickness, both before and after annealing at $\SI{250}{\degreeCelsius}$. The lines represent the values obtained from Eqn.~\ref{eqn:rashba_field_offset} for $\theta_{eff}=-0.02$ with $h_R=0$ (dashed) and $h_R=1.3h_{ad}$ (solid). (b) Effective Gilbert damping for different Fe thicknesses obtained from the frequency dependence of the FMR linewidth. The linear fit to Eqn.~\ref{eqn:epi-irmn-damping-enhancement} yields intrinsic Gilbert damping of $\alpha_0=0.0044$ and effective spin-mixing conductance $G_{eff}=\SI{1.31e15}{\per\ohm\per\meter\squared}$. (c,~d) FMR measured in the $\SI{2}{\nano\meter}$ Fe bars for $\phi=\SI{45}{\degree}$ before (c) and after (d) annealing.}
	\label{fig:epi-irmn-thickness-dependence}
\end{figure}

%\section{\label{sec:epi-irmn-field-like-torque}Field-like Spin-orbit Torque}
The negative spin-Hall angle of IrMn is confirmed in samples with 2 and $\SI{3}{\nano\meter}$ Fe thicknesses. The $h_y/h_{ad}$ ratios for 20 devices across different Fe thicknesses are shown with orange circles in Fig.~\ref{fig:epi-irmn-thickness-dependence}(a). The spin-Hall angle can be calculated using~\cite{TshitoyanThesis}
\begin{equation}
\label{eqn:effective-sh}
\theta_{IrMn} = \frac{1}{T}\cdot\frac{h_{ad}}{h_{Oe}}\cdot\frac{e\mu_0M_sd_{IrMn}d_{Fe}}{\hbar},
\end{equation}
where $d_{IrMn}$ is the IrMn thickness and $T$ is the Fe / IrMn interface transparency, defined as the proportion of the induced spin-Hall current that is transferred into the ferromagnet. Although we cannot calculate the interface transparency T without the spin diffusion length and the conductivity of IrMn~\cite{Tserkovnyak2002,Zhang2015a}, we can confirm that it is approximately the same for the different Fe thicknesses. The effective spin-mixing conductance $G_{eff}$ directly determining the interface transparency is given by
\begin{equation}
\label{eqn:epi-irmn-damping-enhancement}
G_{eff} = (\alpha - \alpha_0)d_{Fe}\frac{2e^2M_s}{\gamma\hbar^2},
\end{equation}
where $(\alpha - \alpha_0)$ is the Gilbert damping enhancement due to spin-pumping through the interface. In Fig.~\ref{fig:epi-irmn-thickness-dependence}(b) we see that $(\alpha - \alpha_0)$ is inversely proportional to the Fe thickness, therefore $G_{eff}$ is constant and the transparency does not vary substantially across the samples. We denote $\theta_{eff}=T\theta_{IrMn}$ in further discussion. 

Using
\begin{equation}
\label{eqn:oersted-and-rashba}
h_y = h_{Oe} + h_{R},
\end{equation}
Eqn.~\ref{eqn:effective-sh} can be re-written as
\begin{equation}
\label{eqn:rashba_field_offset}
\frac{h_{y}}{h_{ad}}=d_{Fe}\frac{e\mu_0M_sd_{IrMn}}{\hbar\theta_{eff}}+\frac{h_R}{h_{ad}},
\end{equation}
where $h_R$ can be the interfacial Rashba field, the field-like term of the spin-Hall STT, or a combination of both. The solid orange line in Fig.~\ref{fig:epi-irmn-thickness-dependence}(a) shows this $h_y/h_{ad}$ ratio for $h_R=1.3h_{ad}$ and $\theta_{eff}=-0.02$, matching well with the measurements. If we use $h_R=0$, the values of average effective spin-Hall angles for the different Fe thicknesses would have to vary between $-0.22$ ($\SI{2}{\nano\meter}$ Fe) and $-0.03$ ($\SI{4}{\nano\meter}$ Fe) to explain the measurements. The order of magnitude difference is not realistic given the same interface transparency and the same thickness of the IrMn layer. The presence of substantial $h_R$ is further confirmed by annealing the samples at $\SI{250}{\degreeCelsius}$ in Argon atmosphere. For the samples with $\SI{2}{\nano\meter}$ Fe the antisymmetric Lorentzian, therefore also $h_y$ flip the sign (Fig.~\ref{fig:epi-irmn-thickness-dependence}(c, d)). The new $h_y$ is opposite the Oersted field, which indicates presence of negative $h_R$ larger than the Oersted field~\cite{Skinner2014}. Values of $h_y/h_{ad}$ after annealing are plotted with black triangles in Fig.~\ref{fig:epi-irmn-thickness-dependence}(a). We note here that if we used the $V_{sym}/V_{asy}$ ratio to determine the spin-Hall angle, we would mistakenly conclude that after annealing the spin-Hall angle is large and positive in the samples with $\SI{2}{\nano\meter}$ Fe.

%\section{\label{sec:epi-irmn-anisotropy-of-spin-orbit-fields}Anisotropy of the Spin-orbit Fields}
\begin{figure}[t!]
	\includegraphics[width=1\columnwidth,angle=0]{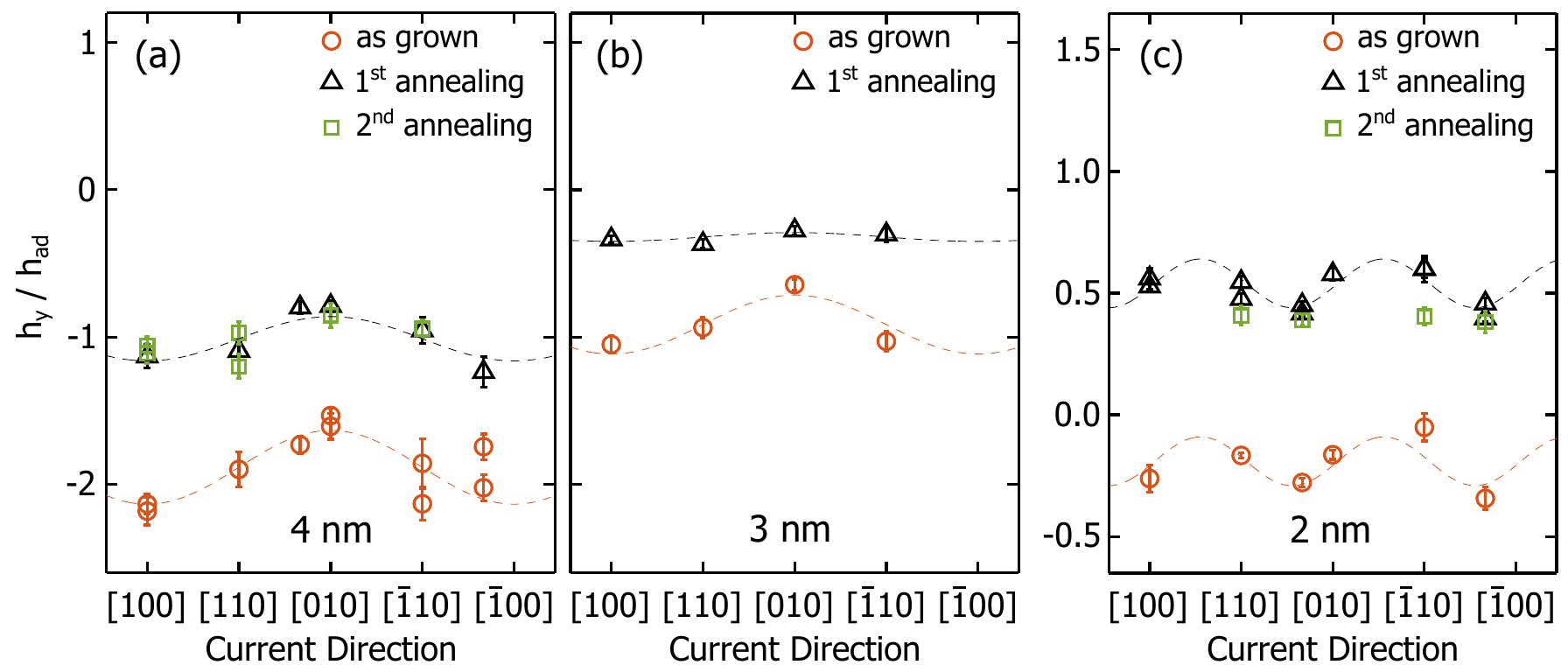}
	\caption{(a~-~c) $h_y/h_{ad}$ ratios measured in bars along different crystallographic directions with different Fe thicknesses. The dashed lines are guides for the eye. Exchange bias for the $\SI{4}{\nano\meter}$ (a) and $\SI{3}{\nano\meter}$ (b) Fe samples is along [100] before annealing, whereas for the $\SI{2}{\nano\meter}$ sample (c) it is between [010] and [$\bar{1}$10]. Annealing in (a) is done with a $\SI{75}{\milli\tesla}$ field along [$\bar{1}$10] ($\mathrm{1^{st}}$) and [100] ($\mathrm{2^{nd}}$), in (b) along [010] and in (c) along [110] ($\mathrm{1^{st}}$) and [$\bar{1}$10] ($\mathrm{2^{nd}}$). In all cases the exchange bias direction is successfully reset after annealing.}
	\label{fig:epi-irmn-torque-anisotropy}
\end{figure}
The $h_y/h_{ad}$ ratios across different devices with the same Fe thickness vary by up to $\SI{30}{\percent}$ (Fig.~\ref{fig:epi-irmn-thickness-dependence}(a)). Interestingly, these variations are not arbitrary and are correlated with the crystallographic direction of current. In Fig.~\ref{fig:epi-irmn-torque-anisotropy}(a~-~c) $h_y/h_{ad}$ is plotted versus the direction of the bar it is measured in (Fig.~\ref{fig:samples}(c)) for each Fe thickness. For the $\SI{3}{\nano\meter}$ and $\SI{4}{\nano\meter}$ Fe, a field along [100] was applied during growth, setting the exchange bias along that direction. We see that $h_y/h_{ad}$ is larger for the current collinear with the exchange bias and smaller when perpendicular to it (Fig.~\ref{fig:epi-irmn-torque-anisotropy}(a, b)). The [100] and [010] directions are otherwise equivalent because of the cubic symmetry of both IrMn and Fe. In the Fe(3) / Pt control sample $h_y/h_{ad}=3.05\pm0.11$ and $3.04\pm0.08$ are found for these bar directions, although the film is grown with the same magnetic field along [100]. This torque anisotropy, however, is not seen in the $\SI{2}{\nano\meter}$ Fe film (Fig.~\ref{fig:epi-irmn-torque-anisotropy}(c)). The small variations of $h_y/h_{ad}$ do not have an apparent uniaxial symmetry. Here, the field was applied along [$\bar{1}$10] during growth, resulting in an exchange bias along a direction between [010] and [$\bar{1}$10]. The misalignment could be due to [$\bar{1}$10] being an in-plane hard axis for Fe, so the field was not enough to fully align the magnetization.

We attempt to control the observed torque anisotropy by resetting the exchange bias. A field of $\SI{75}{\milli\tesla}$ is applied along different directions during annealing, successfully resetting the exchange bias after field cooling. The symmetry of $h_y$, however, is not reset with the changing direction of the exchange bias as seen in Fig.~\ref{fig:epi-irmn-torque-anisotropy}.

%\section{\label{sec:epi-irmn-discussion}Discussion}
The most important observation of our work is perhaps the negative spin-Hall angle. $\mathrm{Ir_{15}Mn_{85}}$ sputtered on Fe at the discussed conditions is chemically disordered~\cite{Kohn2013}. It is expected to have either the theoretically predicted multiple-Q spin density wave structure~\cite{Sakuma2003}, or the experimentally observed cubic-symmetry with moments tilted away by $\SI{45}{\degree}$ from crystal diagonals towards the cube faces~\cite{Kohn2013}. The latter is more likely when the crystal has in-plain strain, which is expected for IrMn grown on Fe. Previous measurements of positive spin-Hall angles have either considered polycrystalline IrMn or chemically ordered $\mathrm{IrMn_3}$ with the triangular magnetic structure. Zhang et al. measured a positive spin-Hall angle in polycrystalline $\mathrm{Ir_{14}Mn_{86}}$~\cite{Zhang2016}, a composition very similar to our $\mathrm{Ir_{15}Mn_{85}}$. Therefore, the exact crystalline and magnetic order is extremely important for spin-Hall effects in antiferromagnets. Similar conclusion was reached in several ab-initio studies for chemically ordered $\mathrm{IrMn_3}$, suggesting that different current directions in the crystal can result in different magnitudes of spin-Hall angles, or even different signs~\cite{Zelezny2017,Zhang2017,Zhang2015}.

The second important observation is the large spin-orbit field with Rashba symmetry opposing the Oersted field that increases after annealing. Previous studies have found that the interfacial Rashba field due to inversion symmetry breaking is mostly reduced by annealing~\cite{Garello2013,Avci2014}. In contrast, if $h_R$ was due to the field-like component of the spin-transfer torque, its increase could be explained by the increased interface transparency after annealing. This would increase both $h_R$ and $h_{ad}$, effectively shifting up the $h_y/h_{ad}$ ratio, in agreement with our observation (Fig.~\ref{fig:epi-irmn-thickness-dependence}(a)). The negative sign of $h_R$ opposite the Oersted field is consistent with the negative spin-Hall angle of IrMn~\cite{Fan2013,Kim2012}.

Lastly, let us consider the anisotropy of the spin-orbit torques correlated with the initial direction of the exchange bias set during growth. Resetting the exchange bias along a different direction does not reset this anisotropy. Therefore, the bulk antiferromagnetic order in IrMn governing the exchange bias is unlikely to be the cause of the anisotropy. Moreover, exchange bias is the largest in the $\SI{2}{\nano\meter}$ Fe sample which does not exhibit torque anisotropy correlated with the the exchange bias. Therefore, the anisotropy is not directly governed by the exchange bias. We believe the anisotropy is instead governed by uncompensated magnetic moments at the Fe / IrMn interface. 

A possible explanation relies on differences of interface transparencies for different current directions. Assume that the magnetic field during growth creates uncompensated moments aligned with the field, which is [100] for 3 and $\SI{4}{\nano\meter}$ Fe samples. Charge current along [100] creates spin-polarization along [010], which scatters at the uncompensated moments resulting in reduced transparency. Charge current along [010], in contrast, induces spin-polarization along [100] that scatters less at the interface, leading to a larger transparency. Hence, the observed larger $h_y/h_{ad}$ for current along [010] compared to [100] (Fig.~\ref{fig:epi-irmn-torque-anisotropy}(a, b)). Field-cooling does not reorient the uncompensated moments, therefore the symmetry does not change. The variations are instead slightly reduced due to the reduction of the uncompensated moments at the elevated temperature. The absence of the anisotropy for the $\SI{2}{\nano\meter}$ Fe sample could be due to the non-saturating field during growth leading to a more random orientation of the uncompensated moments.

%\section{Conclusions and Outlook}
In conclusion, we have observed a negative spin-Hall angle of a few percent in chemically disordered epitaxial $\mathrm{Ir_{15}Mn_{85}}$, in contrast to previous measurements of large positive spin-Hall angles. This highlights the importance of the exact crystalline and magnetic structures for spin-Hall effects in antiferromagnets. A large spin-orbit field with Rashba symmetry opposing the Oersted field is also measured, which increases after thermal annealing. This observation shows that using effective field ratios in spin-torque FMR measurements can lead to wrong values and even a wrong sign for the spin-Hall angle. Lastly, magnitudes of spin-orbit torques depend on the direction of current with respect to the exchange bias set during growth. We believe this is governed by the uncompensated moments at the Fe / IrMn interface and is not directly correlated with the exchange bias or the antiferromagnetic order in bulk IrMn.

%\section{Acknowlegements}
Authors would like to thank T.~Jungwirth for his valuable comments. VT would like to thank the Winton Programme for the Physics of Sustainability and Cambridge Overseas Trust for financial support.

\end{document}